\documentclass[pdflatex,sn-vancouver,Numbered]{sn-jnl}

\usepackage{graphicx}%
\usepackage{multirow}%
\usepackage{amsmath,amssymb,amsfonts}%
\usepackage{amsthm}%
\usepackage{mathrsfs}%
\usepackage[title]{appendix}%
\usepackage{xcolor}%
\usepackage{textcomp}%
\usepackage{manyfoot}%
\usepackage{booktabs}%
\usepackage{algorithm}%
\usepackage{algorithmicx}%
\usepackage{algpseudocode}%
\usepackage{listings}%
\usepackage{lmodern}

\theoremstyle{thmstyleone}%
\theoremstyle{thmstyletwo}%

\theoremstyle{thmstylethree}%

\begin{document}

\title[Materiality and Risk in the Age of Pervasive AI Sensors]{Materiality and Risk in the Age of Pervasive AI Sensors}


\author*[1]{\fnm{Mona} \sur{Sloane}}\email{mona.sloane@virginia.edu}

\author[2]{\fnm{Emanuel} \sur{Moss}}

\author[3]{\fnm{Susan} \sur{Kennedy}}

\author[4]{\fnm{Matthew} \sur{Stewart}}

\author[5]{\fnm{Pete} \sur{Warden}}

\author[6]{\fnm{Brian} \sur{Plancher}}

\author[4]{\fnm{Vijay} \sur{Janapa Reddi}}

\affil*[1]{\orgname{University of Virginia}, \orgaddress{\city{Charlottesville}, \state{VA}, \country{USA}}}

\affil[2]{\orgname{Intel Labs}, \orgaddress{\city{Hillsboro}, \state{OR}, \country{USA}}}

\affil[3]{\orgname{Santa Clara University}, \orgaddress{\city{Santa Clara}, \postcode{95053}, \state{CA}, \country{USA}}}

\affil[4]{\orgname{Harvard University}, \orgaddress{\city{Boston}, \state{MA}, \country{USA}}}

\affil[5]{\orgname{Stanford University}, \orgaddress{\city{Stanford}, \state{CA}, \country{USA}}}

\affil[6]{\orgname{Barnard College, Columbia University}, \orgaddress{\city{New York}, \state{NY}, \country{USA}}}

\abstract{Artificial intelligence (AI) systems connected to  sensor-laden devices are becoming pervasive, which has significant implications for a range of AI risks, including to privacy, the environment, autonomy, and more. There is therefore a growing need for increased accountability around the responsible development and deployment of these technologies. In this paper, we highlight the dimensions of risk associated with AI systems that arise from the material affordances of sensors and their underlying calculative models. We propose a sensor-sensitive framework for diagnosing these risks, complementing existing approaches such as the U.S. NIST AI RMF and the EU AI Act, and discuss its implementation. We conclude by advocating for increased attention to the materiality of algorithmic systems, and of on-device AI sensors in particular, and highlight the need for development of a sensor design paradigm that empowers users and communities and leads to a future of increased fairness, accountability, and transparency.}




\maketitle


\section{Introduction}\label{sec:intro}
Over the past decade, several overlapping, multidisciplinary communities of research and development have emerged to analyze and address the implications of artificial intelligence (AI) systems operating across society, particularly through ethics, engineering, governance, critical academic, and advocacy perspectives (See especially the ACM Conference on Fairness, Accountability, and Transparency, the AAAI/ACM Conference on AI Ethics and Society, and the ACM Conference on Equity and Access in Algorithms, Mechanisms, and Optimization). This work has focused crucial attention on how datasets, algorithms, and machine learning systems deployed at scale produce significant impacts for human rights, equity, already-disadvantaged populations and communities, and society at large~\cite{metcalf2021algorithmic, smuha2021beyond, shelby2023sociotechnical, weidinger2021ethical, selbst2021institutional, birhane2021algorithmic, barocas2016big, noble2018algorithms, eubanks2018automating}. Furthermore, this work has expanded the scope of investigations of these impacts beyond a relatively narrow focus on, e.g., bias in algorithms ~\cite{danks2017algorithmic, mitchell2021algorithmic} or datasets ~\cite{mehrabi2021survey, hazirbas2021towards} to the situated interactions of complex computational systems in society~\cite{ehsan2022algorithmic, metcalf2021algorithmic}. Only recently, this scope has also broadened to include the material dimensions of AI systems, particularly around the environmental impacts of computation in terms of carbon-intensive energy consumption and the geographic footprint of server infrastructure~\cite{dodge2022measuring, bender2021dangers, weidinger2022taxonomy}, as well as e-waste~\cite{bridges2023geographies, kidd_energy_2023}. However, relatively scant attention has been paid to the materiality of AI in terms of data collection devices and particularly sensors that interact with the physical environment (i.e., the sensing of physical phenomena to produce data) and enact algorithmic inferences (e.g., AI-enabled Internet-of-Things (IoT) devices). 

There are significant material dimensions to problems of AI fairness, accountability, and transparency that can be addressed by understanding how ubiquitous, algorithmically-enabled sensors produce risks around bias, equity, privacy, accountability, transparency, and consent. Materiality is a complex term, constituted by both the physical existence of an object, and the social treatment thereof. There are many things that concretely materialize in our social world without taking on physical properties: the perception (or atmosphere) of a space, the potential of an idea, or our social status. Other things assert themselves clearly through the physical presence: bridges, buildings, heirloom objects, or people. Because materiality is integrated into social practices and social institutions, it profoundly shapes sociotechnical systems, like AI  \cite{law1995notes, pinch2008technology, lievrouw2006introduction, miller2020materiality}. This becomes obvious in the context of the pervasive data-collection that is facilitated by the material properties of the components of the many technological devices we use in our daily life: cameras, microphones, batteries, and so on.

Today's sensor technology has shrunk the material components needed to turn physical phenomena into data onto a very small footprint. Increasingly small devices are equipped with microphones, cameras, and machine learning abilities, bringing AI to the materiality of the sensor (rather than communicating data with the cloud). At the smallest scale, and at the lowest power of operation, this is referred to as ``TinyML''~\cite{warden2019tinyml,Reddi2022Widening}. The materiality of these new types of sensors is deeply implicated in the pervasive data-collection that underpins AI systems. In this paper, we follow Lievrouw in framing materiality as ``the physical character and existence of objects and artifacts that makes them useful and usable for certain purposes under particular conditions'' ~\cite[p. 25]{lievrouw2014materiality}. We build on this definition to draw attention to the material design of sensors and its impact on how physical phenomena are transformed into data. By doing so, we propose that there are certain risks associated with the designed materiality of sensors that common ethical approaches (for example, technomoral virtues~\cite{vallor2016technology}) or recent AI risk frameworks (for example, the risk management framework (RMF) developed by the U.S. National Institute of Standards and Technology (NIST)~\cite{nist_artificial_2023} or the EU AI Act~\cite{veale2021demystifying}) do not sufficiently attend to. Similarly, past reviews on the impact of AI on sensing technology mostly focus on the technical challenges of model compression at the edge and large-scale IoT architectures. And, while some may mention privacy and security in passing, they pay little attention to the embodied material challenges of such technological shifts~\cite{mukhopadhyay2021artificial,singh2023edge,haick2021artificial,zhang2022artificial,masson2024roadmap,ullo2021advances,merenda2020edge,zhu2023pushing,abadade2023comprehensive,dutta2021tinyml}.

We stipulate that this blind spot in AI ethics and governance primarily stems from ignoring the material \textit{affordances} of technical objects, and sensors specifically. Foundational texts in science and technologies studies define affordances as the properties of objects that ``are compatible with and relevant for people's interactions''~\cite{gaver1991technology} (see also ~\cite{gibson1977theory}). Here, we build on more recent work that positions the affordances of \textit{technologies} as ``mediating between a technology’s features and its outcomes''~\cite{davis2020artifacts} to focus our attention on the properties of sensors that are relevant for data collection (e.g., sensitivity to physical phenomena, on-board processing, storage, and transmission), noting that affordances can both enable and constrain interactions~\cite{kennewell2001using}. While originally framed as the the way in which design features enable and constrain user engagement and social action~\cite{davis2020artifacts}, we propose to understand \emph{materiality} as the material affordances of sensors which deeply affect data ontology (i.e., considerations of what phenomena can and ought to be data-fied), data collection, and data processing--all of which affect how the benefits and risks of AI systems unfold further downstream. We note that the benefits of such material affordances are often clearly identified by those who design and market sensors, but risks are less frequently identified by designers or articulated for customers or the broader public. Drawing on a growing body of work on AI risks, we define AI risks as concrete harms that can be experienced by individuals or communities or that can be afflicted on the environment through the deployment and use of AI systems~\cite{smuha2021beyond, acemoglu2021harms, kusche2024possible, watkins2021governing, bengio2024managing}. 

For illustration, we can consider the way physical properties change the performance of 
sensors for varying skin tones~\cite{roth2009looking, galdino2001standardizing,guo2023calibrating}. For example, the physical properties of the charge-coupled devices (CCDs) inside digital cameras, a paradigmatic sensor, and the algorithms that process their outputs, contribute to how skin tones are rendered in digital files \cite{roth2009looking}, which in turn contribute to computer vision applications like face detection and facial recognition~\cite{buolamwini2018gender}. 

Adopting a material lens and an affordances approach also makes it possible to apprehend the risks that adhere not to just a single sensor, device, or AI system, but instead emerge from their widespread adoption. We stipulate that adoption is driven by how material properties and functionalities combine with \textit{calculative behaviors and calculative models}, i.e., the collectively shared assumptions and practices about both usefulness and economic value that render sensors commercially viable (these include, but are not limited to, cost of energy for production and use, sensor size, supply chain considerations of material components, as well as profit projections and pricing schemes)~\cite{muniesa2007introduction, callon2005qualculation}. Calculative models directly affect sensor price, availability, and ubiquity, making the concept particularly useful for understanding drivers behind the proliferation of sensors.  As we will show, low production cost and particular calculative models of cameras, microphones, and other sensors make them increasingly common in a wide range of contexts. 

The calculative models that drive the increased affordability and accessibility of sensors have led to positive developments. The integration of sensors in home appliances and other consumer goods can improve their utility and performance. For example, there is a small camera in the Keurig K-Supreme Plus Smart coffee maker that detects the type of pod being inserted in order to adjust the water temperature for optimal brewing. Sensors have also enabled beneficial applications such as predictive maintenance for public infrastructure~\cite{engineering2020windturbine} and industry~\cite{restle2002clock}, providing support for safe driving practices~\cite{flores2023tinyml}, and optimizing energy usage in offices and homes~\cite{shah2019review}. Most notably, sensors coupled with edge computing have opened up novel applications that can make progress towards the 17 sustainable development goals (SDGs) outlined in the United Nations’ 2030 Agenda for Sustainable Development~\cite{UNSDGs2015,Prakash23SustainableTinyML}. For example, sensors are being used to optimize agriculture~\cite{mrisho2020accuracy,king2017technology} and aid wildlife conservation efforts~\cite{Solana20Elephant,Temple19Using,Johnson20Google}. 

However, sensors also give rise to significant concerns as their proliferation into the public, professional, and intimate dimensions of our daily routines enables unprecedented data mining and commercialization of once private or anonymous moments and behaviors~\cite{elmqvist2023data, zuboff2018surveillance}. The material affordances of sensors embedded in mobile devices (such as cameras, microphones, gyroscopes, GPS antennae, and inertial measurement units) and home electronics are often opaque and unaccountable in ways that make it difficult for anyone to understand when and what information they might be collecting and analyzing. Additionally, it is well-known that sensor-driven surveillance technologies are more likely to be deployed in already over-surveilled communities and professions~\cite{fernback2013sousveillance, monahan2017regulating, sevignani2017surveillance}, exacerbating disparate AI impacts and inequities~\cite{gilman2018surveillance, parsons2015beyond}.

In this paper, we examine the materiality of AI risk production by paying close attention to how sensors are incorporated into AI systems and vice versa.  We suggest that the material affordances and calculative models of sensors contribute to the growing ubiquity of sensors and the general risks of AI systems. To do so, we first demonstrate how the material affordances and calculative models of sensors co-evolved over the past half-century by laying out the `evolutionary history' of sensors that culminates in today's environment of pervasive AI-enabled sensing. Leading on from this analysis, we build on the two most prominent and widely adopted AI risk management frameworks in the US and in Europe--the RMF developed by the U.S. NIST ~\cite{nist_artificial_2023} and the EU AI Act~\cite{european_commission_ai_2024}--to propose a sensor-sensitive \textit{AI risk }\textit{identification} framework. In a third step, we raise a call for the development of a new sensor design paradigm that addresses the  risks posed by the convergence of ubiquitous sensing and AI technologies, particularly in the context of TinyML.
Overall, the key contribution of our work is to expand attention to the material affordances and calculative models of sensor-based AI systems when engaging in AI risk diagnostics.


\section{Sensors, Everywhere}\label{sec:everywhere}
Sensors are devices that convert ``a physical phenomenon into an electrical signal''~\cite[p. 1]{wilson2004sensor} that can then be used to quantify environmental aspects such as light, heat, and pressure. They are designed to monitor phenomena within and beyond our human perception, such as imagery, movement, sound, or chemical composition. They transform physical phenomena into numerical representations, adding to the ``avalanche'' of numbers~\cite{hacking2015biopower} that make it possible for the world---people, commodities, communities, and nature---to be represented computationally, for behaviors to be analyzed. Crucially, they are the foundation for many of the datasets that undergird powerful AI technologies. Digital cameras are the sensors that produced the datasets that enabled image recognition~\cite{krizhevsky2012imagenet}. Speech recognition would be impossible without the datasets generated using digital microphones~\cite{ardila2019common}. And many more AI applications, from autonomous vehicles to personal athletic trackers, rely on sensor data as well. However,  technical innovations do not gain widespread adoption merely because of their technical superiority or consumer demand. Rather, they shape and are shaped by a wide range of (often competing) social and commercial interests~\cite{mackenzie_how_1985}. To develop a sensor-sensitive approach to AI risk diagnostics, it is key to unpack how the material affordances of sensor technology co-evolved over time in tandem with the calculative models that motivated their spread and uptake.  


\subsection{The `Evolution' of Sensing}
Sensors have distinct material characteristics that generally remain unchanged: sensitivity to the intensity of an input, dynamic ranges within which phenomena can be accurately transformed into meaningful signals, a propensity to produce noise alongside meaningful signals~\cite{wilson2004sensor}. But sensors' material \emph{\textit{affordances}}---the social actions they enable rather than their abstract characteristics~\cite{gibson2014theory}---have changed dramatically over the last few decades, with significant implications for how they are used, and for the overall AI risk landscape. As technology has advanced, sensors have become smaller, more efficient, more sensitive, and more connected, allowing them to be embedded in various environments and objects, thus providing real-time data and insights that were previously unattainable. This `evolution' has not only extended our ability to understand and interact with the world around us but also paved the way for innovations that continue to shape our future. Sensors have evolved from simple analog devices into `intelligent' systems capable of analyzing data and making decisions at the edge. The advent of wireless sensor networks in the late 1990s marked an important milestone, allowing remote collection and analysis of sensor data \cite{buratti2009overview}. However, sensors were still reliant on external processing power. The subsequent rise of smartphones and IoT devices brought sensors out of isolation, interconnecting them through the internet \cite{mainetti2011evolution}. For the first time, distributed networks of sensors could coordinate to achieve larger goals. However, even as billions of sensors flooded the environment with data, their capabilities remained confined to data collection rather than interpretation.

This changed with the emergence of Edge AI sensors and TinyML sensors, which move machine learning out of data centers and closer to the sensor devices themselves. Edge AI sensors communicate with nearby processors---either separate components within the same device that contains the sensor, or a mobile phone or computer wirelessly connected to the sensor device---where machine learning models run locally~\cite{mukhopadhyay2021artificial,singh2023edge}. 
TinyML sensors incorporate machine learning models directly on the sensors based on new advances in small but powerful models that can run directly on such devices---so called TinyML models~\cite{warden2019tinyml,Reddi2022Widening}. In particular, these models can run directly on the microcontrollers commonly found on physical sensor hardware. This allows for real-time data processing and analysis at the sensor level, without needing to transmit data to a separate computing device.
Furthermore, TinyML sensors usually limit the information shared from the sensor device to only application-specific instructions, e.g., whether or not a person is standing in front of the device, rather than an entire video data stream that could then be put to many purposes off-device~\cite{warden2019tinyml,warden23MLSensors}. Rather than indiscriminately streaming all collected data to cloud servers, Edge AI and TinyML sensors interpret their environment, surface insights, and make decisions not in distant server centers, but at the edge, i.e., on the device or a nearby connected device. 

\begin{figure}[!b]
    \centering
    \includegraphics[width=\textwidth]{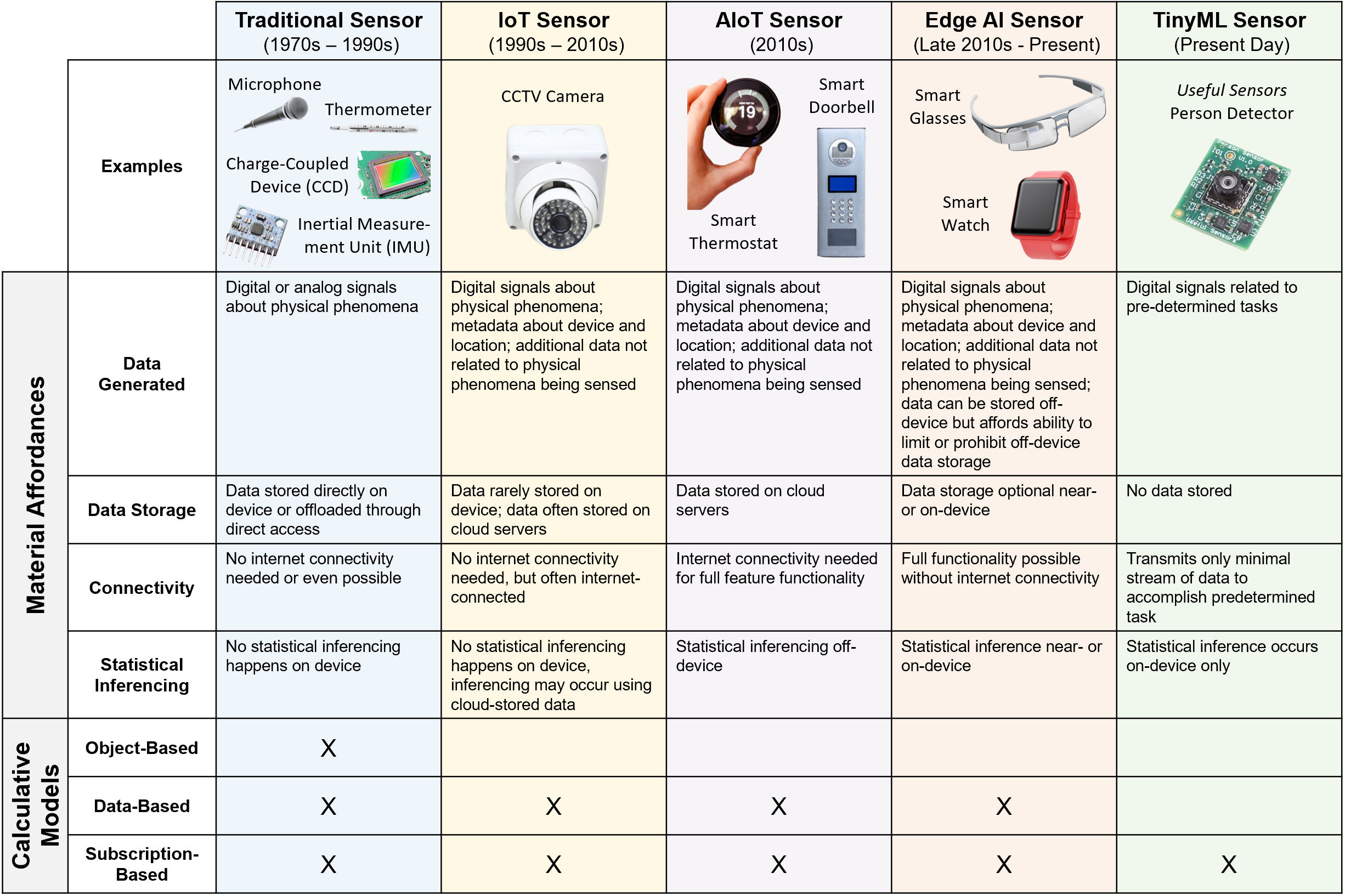} \vspace{-10pt}
    \caption{Timeline of sensor evolution: from passive analog detectors to intelligent IoT and ML-enabled systems. Examples of devices, their material affordances, and their calculative models are also included.}
    \label{fig:evolution}
\end{figure}

Below, we show how successive stages of sensor development, with their evolving material affordances and calculative models (which combine considerations around sensor size, energy use, supply of material components, etc. into profit projections, pricing schemes, and sales strategies), shape the risk profiles of sensors as components of AI systems, providing common examples of each in Fig.~\ref{fig:evolution}, which we then discuss in the following section (Section~\ref{sec:rmfs}). We note that each of the stages of sensor development we identify build on earlier material affordances of sensors while adding additional properties. Accordingly, when thinking about the risks sensors contribute to algorithmic systems we see many of those risks accumulate from one stage to the next.

\begin{itemize}
    \item \textbf{Traditional Sensor~\cite{zhang2004progress,vetelino2017introduction,soloman2009sensors,wilson2004sensor}.} A device that affords the ability to generate data from or about the physical world. The sensor itself may transmit analog or digital signals, and signals produced by the sensor may need post-processing on- or off-device to be useful. Data may be stored on- or off-device, or may be entirely ephemeral. Traditional sensor devices do not require an internet connection to function and typically have no internet connection capability. Data must be purposefully transferred off-device and stored using specific protocols such as SPI or I\textsuperscript{2}C. No statistical inferencing happens on-device.
    
    \item \textbf{IoT Sensor~\cite{li2015internet,rose2015internet,sehrawat2019smart,krishnamurthi2020overview,kocakulak2017overview}.} A device that affords the ability to generate data from or about the physical world, and to access that data in realtime or near-realtime using internet transfer protocols. IoT sensors also afford the ability to collect data not directly related to the physical phenomena they are designed to sense like a list of connected devices, strength of WiFi signals, battery status of connected devices, and other metadata related to the device's performance. IoT sensors may be able to operate without an active internet connection, but most are designed with capabilities that require internet connectivity, particularly to generate or store data. No statistical inferencing happens on-device, and rarely even happens in the cloud. Data transmitted over the internet is (almost) always stored in a centralized location. This type of device acts primarily as an interface to cloud-based data storage systems.

    \item \textbf{AIoT Sensor~\cite{mukhopadhyay2021artificial,vincent2019sensors,haick2021artificial,zhang2022artificial,ullo2021advances,zhu2023pushing}.} A device that affords the ability to conduct cloud-based AI decision-making based on data from or about the physical world. This is done through realtime or near-realtime access to AIoT sensor data using internet transfer protocols, and integrated with AI/ML inferencing techniques in the cloud (and importantly not on or near the device). As such, this type of device requires an active internet connection to operate at full functionality. Importantly, full datastreams generated by the sensor are (temporarily) stored in centralized locations (i.e., cloud or server device). This type of devices acts primarily as an interface to cloud-based decision-making systems powered by AI. 

    \item \textbf{Edge AI Sensor~\cite{singh2023edge,fabre2024near,merenda2020edge,zhu2023pushing,wen2023task,sodhro2019artificial,abadade2023comprehensive,dutta2021tinyml}.} A device that affords the ability to conduct AI decision-making at the location of the device, based on data from or about the physical world. This is done by processing sensor data using machine learning techniques through a combination of on-device and near-device edge processing (e.g., the Apple Watch offloads some computations to a user's cell phone). This data may afford remote access and data reuse instances where it is transmitted off device, and possibly over the internet, in real or near-realtime, in its raw or processed form. Transmitted data may or may not be stored in a centralized location. 
    This type of device extends the cloud-based intelligence of AIoT sensors to the edge.

    \item \textbf{TinyML Sensor~\cite{warden23MLSensors,stewart23MLSensorDatasheets}.} A device that affords the ability accomplish a predetermine task with only the minimal amount of information needed to accomplish that task. For example, a TinyML camera designed for person-detection may only afford the ability to read a single bit from the sensor device (`1' if there is a person within view of the camera and `0' if not). This is done by using machine learning inferencing techniques on data from or about the physical world entirely on-device. Crucially, such sensors do \emph{not} afford the ability to read the raw image data from the sensor device. These devices are typically self-contained and store minimal or no sensed data.
\end{itemize}

It is important to note that these successive stages are `evolutionary' only in a descriptive sense; sensors of each stage remain in production and use today and build upon many of the characteristics of the sensors they were preceded by. But their material affordances are quite different from stage to stage. Where traditional sensors require direct access (physically, over wire, or through radio or infrared signals), IoT and AIoT sensors enable data access from anywhere in the world with an internet connection, data aggregation, and real-time monitoring of many distant locations. Edge AI sensors can afford similar capabilities, but also afford autonomous operation of nearby devices connected to these sensors. TinyML sensors, in contrast, afford the ability to accomplish tasks without needing to support high-bandwidth data flows or process data off-device. Importantly for understanding the AI risks associated with different stages of sensors, each stage has different affordances for misuse (whether intentional or not); traditional sensors and TinyML sensors are more difficult for bad actors to gain unauthorized access to than IoT devices, and do not require the production of datastreams that might make users and passersby vulnerable to privacy breaches or other malicious behavior.


\subsection{Calculative Models and Sensor Development}
\label{sec:calculative}
The wide adoption and evolutionary transformation of sensors can be explained in part by their sheer utility for gathering data about the physical world, and its ascribed usefulness and commercial viability. Additionally, each stage of sensor development adds capabilities that lead to new products and services. But the proliferation of sensors cannot be explained entirely by the technical needs they satisfy. Instead, the integration of sensors into consumer devices, industrial machinery, and civil infrastructure---like any other product---is steered more by the dynamics of calculation for circulation and trade, rather than by technical features. Economic markets are not independent agential entities that are external to social and material life, but rather are collectively organized tools that facilitate the calculation of the value of goods~\cite{callon2005peripheral, muniesa2007introduction, callon2007market}. Calculation bridges quantitative and qualitative aspects in an effort to make goods tradeable.
The practices and cultures of calculation~\cite{hansen2021model} differ by market, but always evolve around certain sets of calculative models, i.e., repeatable ways of calculating an object or good. These calculative models combine cultural knowledges and assumptions about people and consumption (for example, about convenience in the home, stipulating that people will prefer to operate their light switch with voice commands over physically switching on lights) with determinations of the social and behavioral impact of the material affordances of an object or product, costs of manufacture, actual and projected volume of sales, and cumulative profits in the context of the many possible additional factors that affect price, costs, and earnings \cite{besedovsky_financialization_2018, mackenzie_engine_2008}.

There are three interrelated material aspects affecting the evolution of calculative models of sensor technology and their recently accelerated proliferation. The first is a reduction in cost for the production of sensors due to advancements in manufacturing technologies, the development of more affordable material components, and the reduction in cost driven by production at scale~\cite{microsoft_manufacturing_2019}. The second,  relatedly, is the miniaturization of sensors~\cite{huck2022current, ROY201758, RODRIGUEZSAONA2020136}, often considered an innovation- and adoption-driver~\cite{tricoli2017wearable,frazier1995miniaturization,madou2018fundamentals,yang2021miniaturization}. The third one is, again relatedly,  a measurable reduction in energy consumption in sensor deployment. Especially for large networks of sensors, vastly increased energy efficiency of sensors, even for Edge AI and TinyML sensors that run machine learning models at the edge, enhance efficiency by removing the need for cloud services, making sensors capable of real-time data analysis for a wide range of applications where an internet connectivity is impractical or latency restrictions are severe~\cite{jiang2020energy,chen2019energy,rault2014energy,sun2018mems,raha2017towards,schurgers2001energy,Prakash23SustainableTinyML}.

These three interrelated and material aspects  have given rise to three distinct calculative models of sensor technology, with one emerging before the other. While in the 1970s, 1980s, and 1990s, the calculable and tradeable unit was the object of a traditional sensor itself (thermometer, microphone, charge-couple device, or inertial measurement unit), the calculative model changed with the arrival of IoT sensors and sensor networks, starting in the 1990s. Here, the calculable unit is not the sensor or sensor network itself, but the value of the data that it collects. Pricing emerges around the ability of collecting and interpreting that data, for example in the context of  predictive maintenance, energy management, and consumer behavior analysis. A third calculative model later emerged, spanning across all sensor types, focused on subscription-based and service-oriented models as calculative units. Companies are increasingly offering sensors as part of a service package, where the initial cost is low, but users pay a recurring fee for data analysis, cloud storage, and other associated services. 

Today, even mundane appliances have increasing numbers of sensors built into them. There has been an exponential growth in the market size of deployed smart IoT devices, with a total 14.3 billion devices now in use globally in 2022 and a projected 29 billion devices by 2027~\cite{iot_market}.  This proliferation of AI-connected sensors and brings new forms of risk: the likelihood, frequency, and severity of a harm occurring to individuals, communities, or the environment. As sensors variously proliferate, connect to AI services, and embed machine learning capabilities on-device, they shape and reshape how and when algorithmic harms may occur. This has important implications for how we should diagnose AI risk, given the many ways different sensors interact with social worlds and the physical environment, how data is shared between and across devices containing different types of sensors, and how the material affordances of sensors enable the activities that lead to harm.


\section{A Sensor-Sensitive AI Risk Diagnostics Framework}
\label{sec:rmfs}
Increased attention to the harms and risks of AI systems has led to the development of several risk management frameworks for AI (see, e.g., \cite{saif_trustworthy_2020, floridi_capai_2022}) that are variously oriented toward specific business purposes or regulatory conformity. In the U.S., the National Institute of Standards and Technology (NIST) has developed an AI Risk Management Framework (RMF) that is designed as a general purpose framework, applicable across many domains ~\cite{nist_artificial_2023}. The EU has developed a risk framework for AI through the EU AI Act that identifies specific uses of AI as presenting different levels of risk, with some uses prohibited, others high risk enough to be `regulated', and still others requiring varying safety features and transparency mechanisms~\cite{european_commission_ai_2024}. Additional frameworks have been produced by governmental organizations, like the OECD~\cite{oecd_advancing_2023}, or consultancy groups, e.g.,~\cite{baquero_derisking_2020, deloitte_ai_2018}. The NIST AI RMF develops 7 characteristics of what they refer to as ``trustworthy AI systems'', and which are quickly becoming key elements of AI governance. Such systems are Valid and Reliable, Safe, Secure and Resilient, Accountable and Transparent, Explainable and Interpretable, Privacy-Enhanced, and Fair. The NIST AI RMF points out the need for ``balancing each of these characteristics based on the AI system's context of use''~\cite[p. 12]{nist_artificial_2023}. The EU AI Act similarly focuses on contexts of use, by drawing attention to deployments that are explicitly prohibited, those that are regulated as ``high-risk AI systems'', those that have ``fundamental safety and transparency'' concerns, those that require transparency only, and those that are not in the scope of the act~\cite{european_commission_ai_2024}. However, these risk management frameworks set out to provide comprehensive and often too broad guidance on AI risk identification, mitigation, and management, and do not adequately attend to the role pervasive sensing plays in contributing to the risks of AI systems. Therefore, we propose a sensor-sensitive AI risk diagnostics framework that attends to how the material affordances of sensors, and the calculative models that drive their features and deployment patterns, produce specific AI risks within wider AI systems. 

Our approach proceeds from an analysis of the materiality of objects---here, the material affordances of sensors---in contrast to frameworks like NIST's, which proceeds from impacts to users and communities and without attention to the impacts and dynamics of materiality and proliferation. Emerging from our preceding analysis of material affordances and calculative models of sensors, our sensor-sensitive framework for AI risk diagnostics focuses on five key aspects that must be considered as part of assessing AI risk of sensors: 1) calibration, 2) documentation, 3) proprietary data profusion, 4) privacy, and 5) waste. Each of these aspects are intended to be considered \emph{in addition to} other characteristics of an AI system. Overall, our diagnostic ought to be read as complementary to government-mandated AI risk management approaches as we suggest it be used specifically for sensor-intensive AI systems and applications. Our argument is that sensors contribute to the risks of AI systems through their material affordances and calculative models. This contribution is worthy of analysis separate from an analysis focused exclusively on how the data produced by sensors contributes to the risks of AI systems. 


\subsection{Calibration} \label{calibration}
All sensors must be properly calibrated to return validated, reliable, accurate, robust, comparable about the physical world. But the challenge of both achieving and confirming these characteristics is different for various types of sensors~\cite{whitehouse2002calibration}. Traditional sensors are often calibrated before they leave the factory and are engineered either to provide users a means of recalibrating the sensor to a known value (e.g., using standardized weights to adjust a scale or boiling distilled water for a digital thermometer) or including an adjustment curve to adjust observed measures to known measures. Calibration also has a sociotechnical dimension, in that a sensor is calibrated for an intended purpose and can be miscalibrated for other purposes. A thermometer designed for personal health use may need to be calibrated to be most accurate around normal human body temperature, whereas a thermometer designed for industrial use might need to be calibrated differently to be most accurate around the melting point of various substances (e.g., iron, ammonia, or copper). There are numerous instances where sensors have been miscalibrated for their social uses; color film was long miscalibrated to lighter skin tones despite the fact that it was being used for photographing people of a wider range of skin tones and digital photography was developed to emulate color film calibration curves.~\cite{galdino2001standardizing}. Additionally, sensors can become miscalibrated over time, as communities and the environment change; so-called shot-spotter technology requires on-site re-calibration to function properly, particularly when urban geography changes in ways that affect acoustics~\cite{hansen2021gunshot}.

Networked sensors---IoT and AIoT sensors---do not always afford the direct physical access required to conduct or confirm recalibration, so more sophisticated computational techniques may be needed to calibrate already-deployed sensors over a wide area without direct access to the device itself or a reliable reference measurement to calibrate sensor readings~\cite{delaine2019situ}. Edge AI and TinyML sensors also raise validity concerns, as their designed-in autonomy often means calibration cannot be remotely maintained once deployed. The calculative models that underwrite the development and pervasive deployment of sensors also affects their calibration to their environment; designing calibration affordances into sensors adds to their cost, maintaining calibration of a fleet of distributed sensors requires significant organizational infrastructures and costs, and data collected from networked sensors may continue to hold commercial value even if it falls out of calibration. 

Sensors' material affordances shape the data they produce as a condition of its production. Unless they are properly designed for the phenomena they are deployed to sense, the data they produce will be inherently unreliable; oxygen-saturation sensors, for example, must be properly calibrated to a patient's skin tone to work properly~\cite{guo2023calibrating}. Furthermore, because pervasive sensing places sensors into contexts for which they may or may not have been properly calibrated, they are implicated in the reliability and safety of AI systems, as well as their ability to function fairly and manage bias appropriately. This risk is more acute in some use-cases than in others, and so risks associated with the mis/calibration of sensors should be evaluated both in relation to the NIST characteristics of ``trustworthy AI'' and the various levels of risk associated with use cases in the EU AI Act.


\subsection{Documentation}
Traditional sensors have long been developed and deployed through a commitment to safety engineering~\cite{leveson2016engineering}, subject to safety testing, and---like other components---accompanied by standardized datasheets that layout their safe and unsafe usecases~\cite[e.g.]{dewey1998complete}. 
AIoT, Edge AI, and TinyML sensors are similarly accompanied by such datasheets, however these types of sensors afford tight coupling to machine learning, either in the cloud, on device, or nearby. For these tightly coupled systems, datasheets do not provide adequate documentation for key trustworthiness characteristics like interpretability and transparency. As these sensors become more pervasive, their under-documentation makes it increasingly difficult to manage their appropriate deployment, identify cases that require closer scrutiny, and monitor their impacts for potential incidents of harm. 

Existing proposals for documenting Edge AI and TinyML sensors suggest including model cards for their machine learning models~\cite{mitchell2019model, stewart23MLSensorDatasheets}. While this would help mitigate the risks associated with inadequate documentation, effectively managing these risks requires a comprehensive evaluation of whether AI systems---\emph{including the sensors that produce data for them}---are adequately documented. As the use of sensors becomes more widespread, documentation is also crucial for understanding the material affordances of data production and the calculative models under which data may be collected, bundled, and sold. While existing documentation might account for specific data production from a networked sensor, such sensors may afford the production of additional datastreams beyond that which is explicitly documented, producing unanticipated risks~\cite{mitev2020leakypick}. The production and leakage of undocumented data can cause significant harms that may be magnified in specific use cases. Transparency is a key dimension of AI governance, however inadequate documentation of the material affordances of pervasive sensor networks comprises a discrete risk to be identified and managed as well.


\subsection{Proprietary Data Profusion}
The calculative models that underlie a sensor-saturated world tend toward reducing the unit-costs of sensors and ensures their omnipresence. The material devices are becoming cheaper to produce and any excess unit-costs are often rationalized by the calculative models discussed above. This has the result that the amount of data collected, which can be used for providing services (for free or on a subscription plan) \textit{as well as} sold on a secondary market through data brokers~\cite{anthes_data_2015}, is always maximized. Such a model leads observers to conclude that the scale of data collection and aggregation often exceeds individuals' expectations or ability to control~\cite{moyopo2023quantifying, crain2018limits}. 
Given the shift in the calculative models attached to sensors, particularly the perceived need to recoup per unit costs, there is increasing pressures for private- and public-sector organizations to keep data proprietary. This constitutes a risk that the material affordances and calculative models of sensors contributes to, which in turn makes AI systems more risky.

Data profusion brought by sensors has the potential to amplify the divergence between data \textit{quantity} and \textit{quality} and its associated risks. Calculative models are oriented towards the deployment of low-cost sensors which, unlike industry grade sensors, are more susceptible to factors that will result in incomplete or inaccurate data \cite{teh2020sensor}. Sensor data quality can also suffer due to variations in hardware manufacturing, sensor drift, or the state of the battery---as sensor data tends to becomes less reliable as the battery nears the end of its lifespan \cite{ye2016detecting}. Critiques of ``big data'' practices underscore the failure of datasets to offer an objective, accurate, and comprehensive portrayal of real-world phenomena~\cite{d2023data, o2017weapons, crawford2013hidden}. Given that policy-making decisions and resource allocation often rely on quantifiable information, the invisibility or inaccurate portrayal of specific individuals or phenomena within datasets contributes to their marginalization. An increase in dirty data generated in a sensor-saturated world risks perpetuating these issues. Moreover, data profusion propelled by sensors might exacerbate the risks associated with poor data quality as it will create an overwhelming demand for data cleaning. This is a time-consuming process which requires a domain expert to be done effectively and cannot be easily scaled to meet demand \cite{ridzuan2019review}.


\subsection{Privacy}
The materiality of sensors and the scale of their deployment present a unique set of challenges with respect to privacy that are worthy of explicit attention. The physical characteristics of sensors in terms of their small size allows them to be inconspicuously integrated into one’s surroundings in novel and unexpected ways. This presents a formidable obstacle to safeguarding privacy through the mechanisms of notice and consent. Individuals who are unaware of sensors may be involuntarily subjected to data collection and algorithmic processing. But even if individuals were notified of sensors, the abundance of these devices would render the practice of consent untenable. The time and attention required to provide informed consent in every case would exceed an individual’s finite capacities~\cite{scott2019trouble,froomkin2019big}. 

The NIST AI RMF notes how AI systems can present new privacy risks such as enabling inferences that jeopardize de-identification efforts. This risk will become especially salient in a sensor saturated world, as an increase in the volume and breadth of data is positively correlated with the ability to draw such inferences. Moreover, the process of sensor fusion, where data from multiple sensors is combined, enables inferences that would otherwise not be possible from a single data stream~\cite{elmenreich2002introduction}. Pervasive sensing, therefore, provides material affordances for expanding the potential to draw inferences and cross-reference data not only contributes to re-identification risks of de-identified data~\cite{sweeney2000simple}, but also threatens an individual’s ability to exercise control over their data in at least two respects. First, the potential to draw inferences means that individuals may not fully understand either the fact of data collection (i.e., the fact that data was being collected by a sensor at all) or the implications of data collection, casting doubt on the informed nature of their consent. Second, an individual’s decision to opt out of data collection may be rendered futile, as the decisions of others to permit data collection can enable inferences that inadvertently implicate those who seek to abstain, eroding the notion of individual agency in data sharing~\cite{barocas2014big,ding2019survey}. 


\subsection{Waste}
The environmental impact of AI is addressed in risk management frameworks such as the NIST AI RMF, which explicitly identifies risks of harm to the environment within its definition of ``safe'' AI systems. This framework briefly references ``conditions [under which] the environment is endangered'' ~\cite[p. 14]{nist_artificial_2023}, although it lacks operational guidance for assessing these risks. While broader discussions about the environmental costs of computing technologies are more developed elsewhere~\cite{dhar2020carbon,wu2022sustainable,van2021sustainable,lannelongue2021green}, these discussions predominantly focus on the operational activities associated with product use, such as the energy consumption required for training and running inference on an ML model. We contend that this framing does not fully capture the risks posed by the material production and disposal of sensors~\cite{cooper2021dog, bridges2020material, bridges2023geographies}. Below, we detail the unique environmental challenges associated with sensors to illustrate why waste is a distinct and critical aspect of risk for these devices.

First, the environmental risks of sensors differ significantly from other AI systems owing to their reduced carbon output associated with operational activities. The operation of battery-powered sensors, even at a massive scale, can be expected to consume significantly less energy than other computing technologies~\cite{Prakash23SustainableTinyML}. Instead, the environmental impact of these devices lies primarily in the manufacturing and disposal phases, where carbon emissions and other risks are tied to the supply chain and end-of-life processing~\cite{gupta2021chasing}. The biggest contributing factor to their embodied footprint is the batteries that power them~\cite{Prakash23SustainableTinyML}. Coin cell batteries, for example, pose significant challenges due to their small size, short lifespan, and toxic components (e.g., lithium, mercury, cadmium). Since these batteries might not be easily recyclable or biodegradable, they may accumulate in landfills or, worse yet, contribute to pollution and environmental hazards as a result of their improper disposal. The second largest factor of sensors’ carbon footprint is the sensor components themselves, which often rely on rare earth elements and carry high extraction costs including habitat destruction, water pollution, and other ecological impacts. Recent research has shown that significant improvements can be made with regards to reducing the waste impact of sensors by way of integrating sustainable materials into sensor design~\cite{ozer2024bendable}.

Although these environmental risks may resemble those associated with other consumer electronics, the distributed nature of sensors nevertheless raises distinct concerns. Sensor deployments in agriculture, wildlife or environmental monitoring can involve hundreds or even thousands of devices spread across large or remote areas. Unlike consumer electronics, which are typically consolidated in urban areas with access to recycling programs and associated infrastructure, retrieving and responsibly disposing of sensors may be impractical. Thus, there is a risk that sensors will contribute to ``stranded e-waste'' that accumulates in inaccessible or remote environments. Furthermore, sensor devices often operate invisibly or autonomously due to their miniaturization and edge computing design. This materiality of sensors contributes to the risk of ``invisible pollution,'' where discarded devices contribute to environmental harm in ways that may go unnoticed by regulators or consumers. Unlike centralized computing technologies which are easier to oversee and optimize for sustainability, the decentralized and distributed nature of sensors presents novel challenges for managing the associated environmental risks. 

Additionally, sensor-based AI systems are often viewed as environmentally-friendly compared to other AI systems due to their low energy requirements during operation. When deployed towards sustainable aims, their overall carbon footprint may even be net-negative~\cite{Prakash23SustainableTinyML}. However, the energy-efficiency of sensors can, counterintuitively, contribute to their broader environmental risk. When sensors are seen as energy-efficient, this framing can obscure the environmental impact associated with their end-of-life processing and disposal. Moreover, when applied to sensors, a theory in economics known as Jevons' Paradox suggests that increased efficiency can lower the perceived environmental cost, encouraging widespread adoption and ultimately increasing resource consumption~\cite{sorrell2009jevons}. 

Finally, the low cost and scalability of sensors further compound these risks. While the affordability of sensors is often celebrated as an innovation that can enable their global adoption, it may inadvertently promote an overreliance on sensor-based solutions. Given the low-cost and accessibility of sensors, they may be preferred over non-technological alternatives such as policy reforms or community-based interventions, even when the latter have lower environmental costs. The calculative models and materiality of sensors leads to increased affordability, energy efficiency and smaller size devices, spurring their widespread application and adoption. While sensors hold promise for supporting sustainability efforts, the distributed nature of their deployment and their potential to have a paradoxical effect on resource consumption present distinct environmental risks that must not be overlooked. 


\subsection{Implementation}
To implement the above risk diagnostics framework, efforts must be made to \emph{map}, \emph{measure}, and \emph{mitigate} the ways in which sensors contribute to the overall risk of an AI system. \emph{Mapping} risks consists of identifying how the specific sensors employed as part of that system might become miscalibrated or be poorly calibrated, how they might need to be documented, what data they produce, how they contribute to or undermine privacy, and what forms of waste their production and use represents. Particular attention ought to be paid to the material affordances of these sensors, to identify what features or components of the sensors are associated with each risk category. \emph{Measuring} risks consists of identifying the severity and scope of each risk category. A privacy risk might be slight, with respect to any one user, yet have widespread implications for thousands or even millions of users. Conversely, it may be severe but occur rarely. Both dimensions must be evaluated to accurately assess the overall risks. Special attention should be given to the calculative models at play and how they shape the severity and scope of identified risks. \emph{Mitigating} risks consists of identifying the roles that material affordances and calculative models play in each identified risk and exploring how these elements could be modified. For instance, if a particular privacy risk is given broader scope due to a data retention policy that enables data brokers to buy and sell user data, implementing a data deletion policy could help reduce some of that risk. Additionally, alternate designs may be developed to partially mitigate the risk of waste, and active monitoring of reference devices might be undertaken to mitigate the risk of miscalibration over time.


\section{Discussion and Future Work}\label{sec:discuss}

As sensors become interwoven into the fabric of everyday life, they may fade into the background of conscious experience. Nevertheless, these devices are capable of exerting powerful influences over individuals' choices and behavior. AI sensor-based systems can subtly or overtly exert power, whether through persuasion or coercion \cite{verbeek2009ambient}. Activity trackers, for instance, employ a subtle approach by using data-driven feedback to encourage users to exercise more. In contrast, a seat belt sensor that locks the car ignition until the driver buckles up represents a more coercive form of influence. The evolving material affordances of sensors-–-particularly their shrinking size and enhanced processing capabilities-–-will allow for their integration into devices that are ever more intimately linked to individuals’ daily routines ~\cite{nafus2016quantified}. This reality, combined with the calculative models driving the creation of a sensor-saturated world, suggests that individuals will likely be increasingly subjected to technological influences.

AI-sensor based systems can offer significant utility which make them attractive to consumers. For instance, despite users’ awareness of the privacy and security risks associated with IoT devices, research reveals a prevalent ‘I want it anyway’ attitude among users \cite{wang2018want}. However, as AI sensor-based systems become more pervasive, users may reassess their willingness to accept the associated risks. One study indicates that users are less inclined to embrace data collection if they perceive it as excessive within the broader context of their lives: ``Smart home technology was often viewed as a further invasion of, or threat to, privacy in a society where already too much personal information is collected and stored''~\cite[p. 369]{balta2013social}. Notably, individuals’ attitudes are context-sensitive; their perception of risk associated with a particular IoT application is intertwined with their views on the broader proliferation of AI sensor-based systems. 

The proliferation of sensors gives rise to new dimensions of risk that are capable of being felt by the general public. Yet existing approaches to AI ethics and governance overlook these risks owing to their narrow focus on specific applications of AI. For example, the NIST RMF for Information Systems and Organizations–which pertains to information systems ranging from cloud-based systems to IoT devices–explicitly avoids considerations related to material affordances of technological objects like sensors~\cite{joint_task_force_transformation_initiative_risk_2018}. 

Similarly, the EU AI Act focuses on regulating AI systems deemed high-risk owing to their use-case, but overlooks broader risks associated with the proliferation of sensors which contributes to proprietary data profusion. The EU Data Act~\cite{european_commission_ai_2024} represents a notable effort to address this gap, aiming to promote equity by allowing individuals and businesses to access data generated by their use of AI sensor-based systems. This recent development in the EU regulatory landscape demonstrates that piecemeal progress is possible. However, a more unified approach to risk management can be achieved by adopting a lens of analysis grounded in both the material affordances and calculative models of sensors. This perspective allows for a comprehensive understanding of the risks associated with AI sensor-based systems by considering the roles that their technical and physical properties play in shaping their impacts.

Beyond risk management, further work is necessary to support efforts toward the responsible design, development, and deployment of sensors. This includes technical teams developing sensors to incorporate interdisciplinary collaborations with social scientists and behaviorists~\cite{rahwan2019machine} into their work, and potentially leads to harnessing the potential benefits of sensors (in part through attention to their material affordances)~\cite{soltoggio2024collective}, particularly for decentralizing power by widening access to TinyML. With sufficient sensor- and AI-literacy, individuals and communities may be able to build sensor-driven AI systems that genuinely benefit them, under terms they define themselves. These could include entirely private and closed health monitoring systems, weather and crop monitoring employed in farming communities, or sensor deployment for citizen research projects focused on environmental justice or other community concerns. 

To animate and harness these benefits alongside strengthened sensor governance through the proposed sensor-sensitive AI diagnostics framework, it is essential to meaningfully engage stakeholder communities to contribute to the creation of inclusive guidelines and best practices. This ensures that the deployment of sensor technologies considers a broad spectrum of perspectives and needs, balancing technological advancement with societal wellbeing~\cite{ada_lovelace_institute_algorithmic_2022, metcalf_relationship_2022}. The focus on material affordances of sensors demonstrated above is just as well-suited to the exploration of how sensors might benefit such communities, on their terms, as it is to the enumeration and analysis of risks.

A range of different stakeholders, all of whom are experts in various applied fields as well as dimensions of risk discussed above, should focus on collaborating to create a sensing paradigm that is aiming to alleviate the negative impacts stemming from the material affordances and calculative models prevalent in today's sensor ecosystem, and developing community-driven approaches to sensor and data use. Additionally, they should  work on scoping transparency in ways that are relevant to the lived experience of interacting with sensors in order to promote the creation of transparent systems that make it easy for users to understand how their data is being used and for what purpose. Incorporating methodologies like the Machine Learning Technology Readiness Levels (MLTRL) framework can provide a structured approach to ensuring these systems are robust, reliable, and responsible from development through deployment \cite{lavin2022technology}.

Additionally, the Machine Learning Sensor paradigm~\cite{warden23MLSensors} can provide a technical frame for risk-aware and community-driven sensor and data use. This approach suggests that sensors should process all data internally and transmit only abstracted, high-level data through a streamlined interface. This adheres to the principle of data minimization, ensuring that raw data remains exclusively accessible to the onboard sensor processor. This architecture not only enhances system self-containment, thereby improving auditability and accessibility, but also empowers users by maintaining control over their raw data, reducing the likelihood of unwarranted data exploitation by commercial and governmental entities. The proliferation of such a paradigm, or the development of alternatives, will be critical in ensuring that responsibility is a core design principle for future sensor systems. While substantial additional work needs to be done to adequately address the risks posed by AI sensor-based systems~\cite{huckelberry2024tinyml}, this approach provides a strong starting point. Parallel efforts can also use the lens of calculative models to intervene in how sensor-focused AI is underwritten financially. Taxation and increased regulatory scrutiny of such devices can shift the development logics away from integrating sensors that facilitate unconstrained data collection in every device.

These efforts can be combined with a focus on interpreting the societal impact of these technologies, advocating for the rights of affected communities, and helping to draft robust regulatory frameworks that govern the ethical use of sensor data. Sociotechnical approaches like this can also play a vital role in AI design~\cite{baxter2011socio, bauer2009designing}, particularly in the public sector~\cite{leslie2019understanding, abbas2021socio}, as well as fostering global public awareness and education~\cite{plancher2024tinyml4d}, ensuring that the implications of sensor technology are widely understood.

\section{Conclusion}\label{sec:conc}

This paper highlights the dimensions of risk associated with AI systems that arise from the material affordances of sensors and their underlying calculative models. It proposes a sensor-sensitive framework for diagnosing these risks, complementing existing approaches such as the NIST AI RMF and the EU AI Act. A key advantage of this framework is its emphasis on a broader range of stakeholders involved in risk diagnostics and management. While guidelines like the NIST AI RMF focus primarily on actors throughout the AI lifecycle, and the EU AI Act primarily applies to providers and deployers of AI systems, the sensor-sensitive approach brings attention to often-overlooked stakeholders. For example, actors involved in the manufacturing, production, and evaluation of sensors have a critical role to play in mitigating risks related to calibration, documentation, and waste management. Additionally, the risk of proprietary data profusion illustrates how economic policy and regulation can complement technical solutions. Thus, the sensor-sensitive approach fills a gap in the existing AI ethics and governance discourse by considering how sensors contribute to the risks of AI systems and implicate key actors beyond those directly involved in the AI lifecycle.

This paper calls for urgent attention to be directed toward developing responsible sensor architectures and regulatory frameworks concerning sensor development and associated data usage. While some recent work provides a commendable starting point, much remains to be done in this evolving field. Furthermore, engaging stakeholders and communities is essential to fully harness the benefits of sensor technologies within a new sensing paradigm. This is particularly important as the risks identified are not exhaustive and may evolve due to changes in the material affordances of sensors, shifts in calculative models, or ongoing AI innovations leading to new use cases.

\section{{Author Contributions}\label{sec:author contributions}
}
M.P.S. and V.J.R. organized the exploratory seminar that led to this paper. All authors developed the problem statement. M.S. developed the theoretical approach. M.S., E.M., S.K., B.P., M.P.S., and V.J.R. designed and executed the analytical approach. S.K., M.S., and E.M. conducted the policy analysis. B.P., M.P.S., V.J.R. and P.W. produced the technical framework and historical analysis. M.S., E.M., S.K., M.P.S., B.P. and V.J.R. wrote the manuscript with input from all authors. B.P. managed layout, formatting, and figure design. M.S. managed the authorship, submission, and revision process. 

\section{Competing Interests Statement \label{sec:comp interests}
}
Pete Warden is a founder and major shareholder of Useful Sensors Inc, which works on privacy-preserving sensor technology. No other authors declare competing interests. 


\end{document}